\documentclass[preprint,aps, amsmath,amssymb]{revtex4-1}
\newcommand{\be}{\begin{equation}}
\newcommand{\ee}{\end{equation}}
\newcommand{\bea}{\begin{eqnarray}}
\newcommand{\eea}{\end{eqnarray}}
\newcommand{\ba}{\begin{array}}
\newcommand{\ea}{\end{array}}

\usepackage{graphicx}
\usepackage{hyperref}
\usepackage{epstopdf}
\usepackage{amsmath}
\usepackage [autostyle, english = american]{csquotes}
\MakeOuterQuote{"}

\begin{document}

\title{	
Diffractive $\rho$ and $\phi$ production at HERA using a holographic AdS/QCD light-front meson wavefunction 
}

\author{Mohammad Ahmady}
\email{mahmady@mta.ca}
\affiliation{\small Department of Physics, Mount Allison University, \mbox{Sackville, New Brunswick, Canada, E4L 1E6}}

\author{Ruben Sandapen}
\email{ruben.sandapen@acadiau.ca}
\affiliation{\small Department of Physics, Acadia University, Wolfville, Nova-Scotia, Canada, B4P 2R6}
\affiliation{\small Department of Physics, Mount Allison University, \mbox{Sackville, New Brunswick, Canada, E4L 1E6}}

\author{Neetika Sharma}
\email{neetika@iisermohali.ac.in}
\affiliation{ 
Department of Physical Sciences,\\
Indian Institute of Science Education and Research Mohali,\\
S.A.S. Nagar, Mohali-140306, Punjab, India.}

\begin{abstract} 
We use an anti-de Sitter/Quantum Chromodynamics (AdS/QCD) holographic light-front wavefunction for the $\rho$ and $\phi$ mesons, in conjunction with the Color Glass Condensate (CGC) dipole cross-section whose parameters are fitted to the most recent 2015 high precision HERA data on inclusive Deep Inelastic Scattering (DIS), in order to predict the cross-sections for diffractive $\rho$ and  $\phi$ electroproduction. Our results suggest that the holographic meson light-front wavefunction is able to give a  simultaneous description of $\rho$ and $\phi$ production data provided we use a set of light quark masses with  $m_{u,d} < m_{s} \approx 0.14$ GeV.

\end{abstract}
\maketitle


\section{Introduction}
\label{Introduction}

We use the QCD colour dipole model \cite{Nikolaev:1990ja,Mueller:1994jq} together with a non-perturbative  holographic meson light-front wavefunction \cite{Brodsky:2014yha} to predict the cross-sections for diffractive $\rho$ and $\phi$ electroproduction measured at the HERA collider \cite{Adloff:1999kg,Aid:1996bs,Breitweg:1997ed,Chekanov:2005cqa,Aaron:2009xp,Chekanov:2007zr}.  In Ref. \cite{Forshaw:2012im}, successful predictions were obtained for diffractive $\rho$ production using the holographic wavefunction for the $\rho$ and the CGC dipole cross-section \cite{Watt:2007nr} whose parameters were fitted to the 2001 HERA DIS structure function data \cite{Chekanov:2001qu,Adloff:2000qk}. In 2015, the latest high precision combined HERA data on inclusive DIS were released \cite{Abramowicz:2015mha}. This definitive  DIS data set supersedes the earlier ones and is one of the major legacies of the HERA collider. We shall use these new data here to update the parameters of the CGC dipole cross-section and thus repeat the predictions of Ref. \cite{Forshaw:2012im}. We shall also extend our predictions to diffractive $\phi$ production, thereby testing the holographic wavefunction for the heavier $\phi$ meson.  

The holographic meson wavefunction is predicted in holographic light-front  QCD proposed by Brodsky and de T\'eramond \cite{deTeramond:2005su,Brodsky:2006uqa,deTeramond:2008ht}.  A recent review of holographic light-front QCD can be found in Ref. \cite{Brodsky:2014yha}. In a semiclassical approximation of light-front QCD with massless quarks,  there is an exact correspondance between the light-front Schr\"odinger equation for QCD bound states in physical spacetime and the equation of motion of spin-$J$ modes in the higher dimensional AdS space. A dilaton field breaking the conformal symmetry of AdS space then dictates the form of the confining potential in physical spacetime. A phenomenologically successful choice is a dilaton which is quadratic in the fifth dimension of AdS space and this maps onto a light-front harmonic oscillator in physical spacetime. Remarkably, group theoretical arguments based on the underlying conformality of the classical Lagrangian of QCD reveal that the light-front harmonic potential is unique \cite{Brodsky:2013ar}.

A single mass scale, $\kappa$, appears in the quadratic dilaton field and thus in the light-front harmonic oscillator in physical spacetime.  The holographic light-front Schr\"odinger equation can then be solved to predict the meson mass spectrum. The latter has a string model Regge form as is observed experimentally.  The parameter $\kappa$ can then be fixed to fit the observed slopes of the Regge trajectories for the various meson families. It is found that for all light mesons, $\kappa \approx 0.5$ GeV \cite{Brodsky:2014yha}. Furthermore, the pion is predicted to be massless, consistent with chiral symmetry. 
  
Accounting for non-zero quark masses goes beyond the AdS/QCD correspondence and in Ref. \cite{Brodsky:2008pg}, Brodsky and de T\'eramond propose an ansatz for including small (on a hadronic scale) quark masses. The key observation is that the evolution variable for the momentum space light-front wavefunction is the quark-antiquark invariant mass and this can be appropriately modified to account for non-zero quark masses. With the modified holographic wavefunction, the shift in meson masses can be computed  as a first order perturbation.  For the pion (and kaon), the mass shift is equal to the meson's physical mass.  This allows the light quark masses to be fixed for a given $\kappa$. Ref. \cite{Brodsky:2014yha} reports $m_{u,d}=0.046$ GeV and $m_s=0.357$ GeV with $\kappa=0.54$ GeV. The quark masses in holographic light-front QCD are thus effective quark masses, between current and constituent quark masses and they vanish in the chiral limit \cite{Brodsky:2014yha}. Once $\kappa$ and the quark masses are fixed, the holographic meson wavefunction comes with no free parameters.

 In the dipole model, the quark mass acts as an infrared regulator and thus reflects confinement. In practice, its value is chosen to fit inclusive DIS data. The typical value of $0.14$ GeV, which coincides with the pion mass, was  used in early extractions of the dipole cross-section \cite{Forshaw:2004vv,Kowalski:2006hc,Marquet:2007qa} from the inclusive DIS data. It is worth noting that the predictions in Ref. \cite{Forshaw:2012im} were generated using a light quark mass of $0.14$ GeV, i.e, consistent with the fact the fitted parameters of the CGC dipole cross-section used in Ref. \cite{Forshaw:2012im} were obtained using that same light quark mass. The most recent extractions of the dipole cross-section were performed using the 2010 HERA DIS data \cite{Aaron:2009aa} in Refs. \cite{Rezaeian:2013tka,Rezaeian:2012ji}. These authors found that the best fits are obtained using current quark masses $m_{u,d,s} \approx 10^{-3}$ GeV. The preference of the DIS data for lower light quark masses was also noted in Ref. \cite{Watt:2007nr} although the effective quark masses $m_{u,d,s}=0.14$ GeV also gave acceptable fits to the 2001 DIS structure function data. In the recent paper \cite{Contreras:2015joa}, using a new dipole model, both the current quark masses and the effective quark masses $m_{u,d,s}=0.14$ GeV are found to give equally good fits to the 2010 DIS structure function  data \cite{Aaron:2009aa}. In all cases, SU(3) flavour symmetry is assumed.

  We shall start by predicting the vector and tensor coupling constants of the $\rho$ and $\phi$ mesons using their holographic wavefunctions. The vector coupling is also referred to as the decay constant since it is related to the measured electronic decay width. On the other hand, the  (scale-dependent) tensor coupling is not extracted from experiment but non perturbative methods like lattice QCD and QCD Sum Rules are able to predict this coupling at a definite scale. We shall find that we are able to achieve optimal agreement with the decay width data by taking $m_{u,d,s} \lesssim 0.14$ GeV. This upper limit coincides with the light quark mass used in earlier dipole model studies\cite{Forshaw:2004vv,Kowalski:2006hc,Marquet:2007qa}. We are thus led to depart slightly from Ref. \cite{Brodsky:2014yha} by considering two additional sets of quark masses with decreasing strength of SU(3) symmetry breaking: $m_{u,d}=0.046~\text{GeV}; m_s=0.14$ GeV and $m_{u,d,s}=0.14$ GeV. In all cases, we use $\kappa=0.54$ GeV. With each set of quark masses,  we  shall refit the parameters of the dipole cross-section  to DIS data and then use the fitted dipole cross-section to predict diffractive $\rho$ and $\phi$ production without any further adjustment of parameters. We shall see that the quark mass set with intermediate SU(3)  symmetry breaking is necessary to describe the data on the ratio of  the $\phi$  to $\rho$ total cross-sections.

 We begin by reviewing the colour dipole model in Section \ref{Dipole model} before discussing the holographic meson wavefunction in Section \ref{Holographic wfn}.  In Section \ref{F2 fit}, we report the results of fitting the dipole cross-section to the new 2015 HERA DIS data. We use the dipole cross-section together with the holographic meson wavefunction to compute diffractive cross-sections for $\rho$ and $\phi$ in Section \ref{Diffractive data}. We conclude in Section \ref{Conclusion}.

\section{The dipole model}
\label{Dipole model}

In the dipole picture, the largeness of the centre-of-mass energy squared, $s$, guarantees that the scattering amplitude for the diffractive process $\gamma^* p \to V p$ factorizes into an overlap of photon and vector meson light-front wavefunctions and a dipole cross-section \cite{Watt:2007nr}:
\bea
 \Im \mbox{m}\, \mathcal{A}_\lambda(s,t;Q^2)  
 &=&   \sum_{h, \bar{h}} \int {\mathrm d}^2 {\mathbf r} \; {\mathrm d} x \; \Psi^{\gamma^*,\lambda}_{h, \bar{h}}(r,x;Q^2)  \Psi^{V,\lambda}_{h, \bar{h}}(r,x)^* e^{-i x \mathrm{r} \cdot \mathbf{\Delta}} \mathcal{N}(x_{\text{m}},\mathrm{r}, \mathbf{\Delta})
\label{amplitude-VMP} 
\eea
where $t=-\mathbf{\Delta}^2$ is the squared momentum transfer at the proton vertex. $\Psi^{\gamma^*,\lambda}_{h, \bar{h}}(r,x;Q^2)$ and $\Psi^{V,\lambda}_{h, \bar{h}}(r,x)$  are the light-front wavefunctions of photon and vector meson respectively while $\mathcal{N}(x_{\text{m}},\mathrm{r},\mathrm{\Delta})$ is the proton-dipole scattering amplitude. The light-front wavefunctions are the probability amplitudes for the virtual photon or vector meson to fluctuate into a $q\bar{q}$ color dipole in a given helicity configuration ($h$ is the helicity of the quark and $\bar{h}$ is the helicity of the antiquark) and they depend on the transverse size $r$ of the $q\bar{q}$ color dipole and on $x$, the fraction of light-front momentum of the photon (or vector meson) carried by the quark. Both wavefunctions are labelled by $\lambda=L,T$ which denotes the polarization of the photon or vector meson. The photon light-front wavefunction is also a function  of the photon's virtuality $Q^2$. The dipole-proton  scattering amplitude  is the amplitude for the elastic scattering of the dipole on the proton and it depends on the photon-proton centre-of-mass energy via the modified Bjorken variable $x_{\mbox{m}}$ where \cite{Rezaeian:2013tka}
\begin{equation}
	x_{\text{m}}=x_{\text{Bj}}\left(1+ \frac{M_V^2}{Q^2} \right)~\text{with}~x_{\text{Bj}}=\frac{Q^2}{W^2}
\;.
\label{Bjorken-x}
\end{equation}
The dipole-proton scattering amplitude contains all the high energy QCD dynamics of the dipole-proton interaction.  It is a universal object, appearing also in the formula for the fully inclusive DIS process: $\gamma^* p \to X$. Indeed, replacing the vector meson by a virtual photon in Eq. \eqref{amplitude-VMP}, we obtain the amplitude for elastic Compton scattering $\gamma^* p \to \gamma^* p$, i.e.
\bea
\left. \Im \mbox{m}\, \mathcal{A}_\lambda(s,t) \right|_{t=0}  
 &=&  s \sum_{h, \bar{h}} \int {\mathrm d}^2 {\mathbf r} \; {\mathrm d} x \; |\Psi^{\gamma^*,\lambda}_{h, \bar{h}}(r,x;Q^2)|^2  \hat{\sigma}(x_{\text{m}}, r) 
\label{amplitude-compton} 
\eea
where we have introduced the dipole cross-section
\begin{equation}
	\hat{\sigma}(x_{\text{m}},r)=\frac{\mathcal{N}(x_{\text{m}},\mathrm{r}, \mathbf{0})}  {s}=\int \mathrm d^2 \mathbf{b}~\mathcal{\tilde{N}}(x_{\text{m}},\mathrm{r}, \mathbf{b}) \;.
	\label{dipole-xsec}
\end{equation}
Via the Optical Theorem, the elastic amplitude given by Eq. \eqref{amplitude-compton} is directly related to the inclusive $\gamma^* p \to X$ total cross-section in DIS: 
\begin{equation}
	\sigma_{\lambda}^{\gamma^* p \to X} = \sum_{h, \bar{h},f} \int {\mathrm d}^2 {\mathbf r} \; {\mathrm d} x \; |\Psi^{\gamma^*,\lambda}_{h, \bar{h}}(r,x;Q^2)|^2  \hat{\sigma}(x_{\text{m}}, r)
	\label{gammapxsec}
\end{equation}
where now  \cite{Rezaeian:2013tka}
\begin{equation}
	x_{\text{m}}=x_{\text{Bj}}\left(1+ \frac{4m_f^2}{Q^2} \right)~\text{with}~x_{\text{Bj}}=\frac{Q^2}{W^2}
\;.
\label{Bjorken-x-DIS}
\end{equation}

This means that one can use the high quality DIS data from HERA to constrain the free parameters of the dipole cross-section section and then use the same dipole cross-section to make predictions for vector meson production and other distinct processes like Deeply Virtual Compton Scattering (DVCS) and Diffractive DIS. This program has been successfully carried by several authors \cite{Forshaw:2006np,Kowalski:2006hc,Marquet:2007qa,Watt:2007nr,Rezaeian:2013tka,Rezaeian:2012ji} hinting very strongly at the universality of the dipole cross-section. 

Note that the high energy factorization in Eqs. \eqref{amplitude-VMP} and \eqref{gammapxsec} holds beyond the validity of perturbation theory, i.e. for all dipole sizes. In practice, the expressions for the photon light-front wavefunctions obtained perturbatively in light-front QED are used for all $r$. To lowest order in $\alpha_{\mbox{em}}$, the perturbative photon wavefunctions are given by \cite{Lepage:1980fj,Dosch:1996ss,Forshaw:2003ki,Kulzinger:1998hw}:
\bea \Psi^{\gamma,L}_{h,\bar{h}}(r,x; Q^2, m_f)  &=& \sqrt{\frac{N_{c}}{4\pi}}\delta_{h,-\bar{h}}e\, e_{f}2 x(1-x) Q \frac{K_{0}(\epsilon r)}{2\pi}\;, 
\label{photonwfL} \\
\Psi^{\gamma,T}_{h,\bar{h}}(r,x; Q^2, m_f) &=& \pm \sqrt{\frac{N_{c}}{2\pi}} e \, e_{f} 
 \big[i e^{ \pm i\theta_{r}} (x \delta_{h\pm,\bar{h}\mp} -  (1-x) \delta_{h\mp,\bar{h}\pm}) \partial_{r}   +  m_{f} \delta_{h\pm,\bar{h}\pm} \big]\frac{K_{0}(\epsilon r)}{2\pi} \label{photonwfT}
\eea
where 
$ \epsilon^{2} = x(1-x)Q^{2} + m_{f}^{2} $ and $r e^{i \theta_{r}}$ is the complex notation for the transverse separation between the quark and anti-quark. As can be seen, at $Q^2 \to 0$ or $x \to (0,1)$, the photon light-front wavefunctions become sensitive to the non-zero quark mass $m_f$ which prevents the modified Bessel function $K_0(\epsilon r)$ from diverging, i.e. the quark mass acts as an infrared regulator. On the other hand, a non-perturbative model  for the meson light-front wavefunction is used and assumed to be valid for all $r$.

To compare with experiment, we compute the  differential cross-section in the forward limit, i.e.

\be
\left. {{\mathrm d} \sigma_\lambda \over dt}\right. \mid_{t=0}  
= \frac{1}{16\pi} 
[ \mathcal{A}_\lambda(s, t=0)]^2
\label{dxsection}
\ee
and we then assume the $t$-dependence to be exponential, i.e.   
\be
\left. {{\mathrm d} \sigma_\lambda \over dt}\right.
= \frac{1}{16\pi} 
[\mathcal{A}_\lambda(s, t=0)]^2  \exp(-B_D t)
\label{dsigmadt}
\ee
where the diffractive slope parameter $B_D$  is given by
\be
B_D = N\left( 14.0 \left(\frac{1~\mathrm{GeV}^2}{Q^2 + M_{V}^2}\right)^{0.2}+1\right)
\label{Bslope}
\ee
with $N=0.55$ GeV$^{-2}$. This parametrization of the diffractive slope  agrees with the most recent ZEUS data for both $\rho$ and $\phi$ production \cite{Chekanov:2007zr}. The most recent H1 data for $\rho$ production \cite{Aaron:2009xp} prefer a somewhat larger value of $B_D$, but with a larger uncertainty. 

Note that Eq. \eqref{dsigmadt} can be rewritten as 
\be
\left. {{\mathrm d} \sigma_\lambda \over dt}\right.
= \frac{1}{16\pi} 
[\Im\mathrm{m} \mathcal{A}_\lambda(s, t=0)]^2 \; (1 + \beta_\lambda^2) \exp(-B_D t)
\label{dsigmadt-re}
\ee
where $\beta_\lambda$ is the ratio of real to imaginary parts of the amplitude. We estimate $\beta_{\lambda}$ in the usual way \cite{Kowalski:2006hc}
\be \beta_\lambda=\tan \left(\frac{\pi}{2} \alpha_\lambda \right) ~~ ~  {\rm with }~~ ~  \alpha_\lambda=\frac{\partial \log |\Im \mathrm{m}\, \mathcal{A}_\lambda|}{\partial \log  \left(1/x\right)} \label{logderivative} \,. \ee 
We calculate the photo-production cross section after integrating Eq. \eqref{dsigmadt} over $t$. This means that the uncertainty in the diffractive slope $B_D$ leads to an uncertainty in the normalization of our predictions for total cross-section. We shall give predictions for the total cross-section $ \sigma = \sigma_T + 0.98 \sigma_L$  to be compared to the HERA data.

\section{Holographic meson wavefunctions}
\label{Holographic wfn}
The vector meson light-front wavefunctions appearing in Eq. \eqref{amplitude-VMP} cannot be computed in perturbation theory. Nevertheless, they can be assumed to have the same spinor and polarization structure as in the photon case, together with an unknown non-perturbative wavefunction \cite{Forshaw:2003ki}. Explicitly, the vector meson light-front wavefunctions can be written as \cite{Forshaw:2012im}
 \be
\Psi^{V,L}_{h,\bar{h}}(r,x) =  \frac{1}{2} \delta_{h,-\bar{h}}  \bigg[ 1 + 
{ m_{f}^{2} -  \nabla_r^{2}  \over x(1-x)M^2_{V} } \bigg] \Psi_L(r,x) \,.
\label{mesonL}
\ee
and
\be \Psi^{V, T}_{h,\bar{h}}(r,x) = \pm \bigg[  i e^{\pm i\theta_{r}}  ( x \delta_{h\pm,\bar{h}\mp} - (1-x)  \delta_{h\mp,\bar{h}\pm})  \partial_{r}+ m_{f}\delta_{h\pm,\bar{h}\pm} \bigg] {\Psi_T(r,x) \over 2 x (1-x)}\,. 
\label{mesonT}
\ee
  
   Various ansatz for the non-perturbative meson wavefunction have been proposed in the literature \cite{Dosch:1996ss,Nemchik:1996cw}, perhaps the most popular one being the so-called Boosted Gaussian (BG) wavefunction \cite{Nemchik:1996cw,Forshaw:2003ki} which has been used in the recent studies in Refs. \cite{Rezaeian:2013tka,Rezaeian:2012ji} to describe simultaneously the cross-section data on diffractive $\rho, \phi$ and $J/\Psi$ production. Ref. \cite{Santos:2014vwa} uses the dipole cross-section extracted in Ref. \cite{Rezaeian:2013tka} with the BG  wavefunction to predict vector meson production in ultrapheripheral collisions at the LHC. In Refs. \cite{Forshaw:2010py,Forshaw:2011yj}, the $\rho$ meson wavefunction is extracted from the data using several dipole models which fit the 2001 DIS structure function data.

   In recent years, new insights about hadronic light-front wavefunctions based on the anti-de Sitter/Conformal Field Theory (AdS/CFT) correspondence have been proposed by Brodsky and de T\'eramond. \cite{deTeramond:2005su,Brodsky:2006uqa,deTeramond:2008ht}. These authors found that  in a semiclassical approximation of light-front QCD with massless quarks, the meson wavefunction can be written as \cite{Brodsky:2014yha} 
   
   \begin{equation}
	\Psi(\zeta, x, \phi)= e^{iL\phi} \mathcal{X}(x) \frac{\phi (\zeta)}{\sqrt{2 \pi \zeta}} 
\label{mesonwf}
\end{equation}
 where the variable $\zeta=\sqrt{x(1-x)} r$ is the transverse separation between the quark and the antiquark at equal light-front time. The transverse wavefunction $\phi(\zeta)$ is a solution of the so-called holographic light-front Schr\"odinger equation:
 \begin{equation}
 	\left(-\frac{d^2}{d\zeta^2} - \frac{1-4L^2}{4\zeta^2} + U(\zeta) \right) \phi(\zeta) = M^2 \phi(\zeta) 
 	\label{holograhicSE}
 \end{equation}
 where $M$ is the mass of the meson and $U(\zeta)$ is the confining potential which at present cannot be computed from first-principle in QCD. On the other hand, making the substitutions $\zeta \to z$ where $z$ being the fifth dimension of AdS space, together with   $L^2 -(2-J)^2 \to (mR)^2$  where $R$ and $m$ are the radius of curvature and mass parameter of AdS space respectively, then Eq. \eqref{holograhicSE} describes the propagation of spin-$J$ string modes in AdS space. In this case, the potential is given by
 \begin{equation}
 	U(z, J)= \frac{1}{2} \varphi^{\prime\prime}(z) + \frac{1}{4} \varphi^{\prime}(z)^2 + \left(\frac{2J-3}{4 z} \right)\varphi^{\prime} (z) 
 \end{equation}
 where $\varphi(z)$ is the dilaton field which breaks the conformal invariance of AdS space. A quadratic dilaton ($\varphi(z)=\kappa^2 z^2$) profile results in a harmonic oscillator potential in physical spacetime:
 \begin{equation}
 	U(\zeta,J)= \kappa^4 \zeta^2 + \kappa^2 (J-1) \;.
 	\label{harmonic-LF}
 \end{equation}
 Brodsky, Dosch and de T\'eramond have shown  that the light-front harmonic potential is unique \cite{Brodsky:2013npa}.  Solving the holographic Schr\"odinger equation with this harmonic potential given by Eq. \eqref{harmonic-LF} yields the meson mass spectrum \cite{Branz:2010ub,Vega:2009zb}
 \begin{equation}
 	M^2= 4\kappa^2 \left(n+\frac{L+J}{2}\right)\;
 	\label{mass-Regge}
 \end{equation}
  with the corresponding normalized eigenfunctions
 \begin{equation}
 	\phi_{n,L}(\zeta)= \kappa^{1+L} \sqrt{\frac{2 n !}{(n+L)!}} \zeta^{1/2+L} \exp{\left(\frac{-\kappa^2 \zeta^2}{2}\right)} L_n^L(x^2 \zeta^2) \;.
 \label{phi-zeta}
 \end{equation}
 
 To completely specify the holographic wavefunction given by Eq. \eqref{mesonwf}, the longitudinal wavefunction $\mathcal{X}(x)$ must be determined. For massless quarks, this is achieved by an exact mapping of the pion  electromagnetic form factors in AdS and in physical spacetime resulting in  \cite{Brodsky:2014yha}.
 \begin{equation}
 	\mathcal{X}(x)=\sqrt{x(1-x)}
 \end{equation}

 For meson families with $J=L+S$, Eq. \eqref{mass-Regge} predicts that the mesons lie on linear Regge trajectories as  is experimentally observed and thus $\kappa$ can be chosen to fit the Regge slope. Ref. \cite{Brodsky:2014yha} reports $\kappa=0.54$ GeV for vector mesons. Eq. \eqref{mass-Regge} also predicts that the pion and kaon (with $n=0, L=0, S=0$) are massless. To account for their physical masses, non-zero light quark masses have to be introduced.  To do so, we follow the prescription of Brodsky and de T\'eramond given in Ref. \cite{Brodsky:2008pg} and which we outline below.

For the ground state mesons with $n=0, L=0$,  Eq. \eqref{mesonwf} becomes
\be 
\Psi (x,\zeta)= \frac{\kappa}{\sqrt{\pi}} \sqrt {x(1-x) }  \exp{\left[-{\kappa^2 \zeta^2 \over 2} \right] } \;.
\label{wavef}
\ee
A two-dimensional Fourier transform to momentum space yields
\be 
\tilde{\Psi} (x,k) \propto  \frac{1}{\sqrt {x(1-x)}}  \exp{\left[-\frac{M^2_{q\bar{q}}}{2\kappa^2}\right] }
\label{wavefk}
\ee
where $M^2_{q\bar{q}}$ is invariant mass of the $q\bar{q}$ pair is given by
\begin{equation}
	M^2_{q\bar{q}}=\frac{k^2}{x(1-x)} \;.
\end{equation}
For non-zero quark masses, this invariant mass becomes
\begin{equation}
	M^2_{q\bar{q}}=\frac{k^2 + m_f^2}{x(1-x)} \;.
	\label{Massqbarq}
\end{equation}
Inserting Eq. \eqref{Massqbarq} in Eq. \eqref{wavefk} and Fourier transforming back to configuration space gives
 \be  \Psi_{\lambda} (x,\zeta) = \mathcal{N}_{\lambda} \sqrt{x (1-x)}  \exp{ \left[ -{ \kappa^2 \zeta^2  \over 2} \right] }
\exp{ \left[ -{m_f^2 \over 2 \kappa^2 x(1-x) } \right]}
\label{hwf}
\ee
where we have introduced a polarization-dependent normalization constant ${\mathcal N}_{\lambda}$. We fix this normalization constant by requiring that
\be
\sum_{h,\bar{h}} \int {\mathrm d}^2 {\mathbf{r}} \, {\mathrm d} x |
\Psi^{V, \lambda} _{h, {\bar h}}(x, r)|^{2} = 1 
\ee
where $\Psi^{V,\lambda}_{h, \bar {h}}(x,r)$ are given by Eqs. \eqref{mesonL} and \eqref{mesonT}.

With the non-zero light quark masses, the meson mass spectrum becomes   \cite{Brodsky:2013npa}
\begin{equation}
	M^2=\Delta M^2 + 4\kappa^2 \left(n+L +\frac{S}{2}\right) 
\label{mass-spectrum}
\end{equation}
 where the mass shift is given by \cite{Brodsky:2014yha}
 \begin{equation}
 \Delta M^2 =\frac{\int_0^1 \mathrm{d} x \exp{\left[-\frac{1}{\kappa^2}\left(\frac{m_f^2}{x(1-x)}\right) \right]}\frac{m_f^2}{x(1-x)}}{\int_0^1 \mathrm{d} x \exp{\left[-\frac{1}{\kappa^2}\left(\frac{m_f^2}{x(1-x)}\right)\right]}} \;.
 \end{equation}
Hence, Eq. \eqref{mass-spectrum} implies that
 \begin{equation}
 	\Delta M = M_{\pi^{\pm}} = 140 ~ \mbox{MeV}
 \end{equation}	 
 which allows to fix the $u$ (and $d$) quark masses for a given $\kappa$. Using $\kappa=0.54$ GeV, Ref. \cite{Brodsky:2014yha} reports $m_{u,d}=0.046$ GeV. To fix the strange quark mass, Ref. \cite{Brodsky:2014yha} uses $\Delta M = M_{K^{\pm}}=494$ MeV together with $m_{u,d}=0.046$ GeV and obtain $m_s=0.357$ GeV. Note that the above Brodsky-de T\'eramond quark mass prescription is expected to be a good approximation only for light quark masses. A possible way to account for heavier quark masses is to determine the longitudinal function $\mathcal{X}(x)$ dynamically \cite{Chabysheva:2012fe}.

\begin{figure}[htbp]
\centering 
\includegraphics[width=14cm,height=14cm]{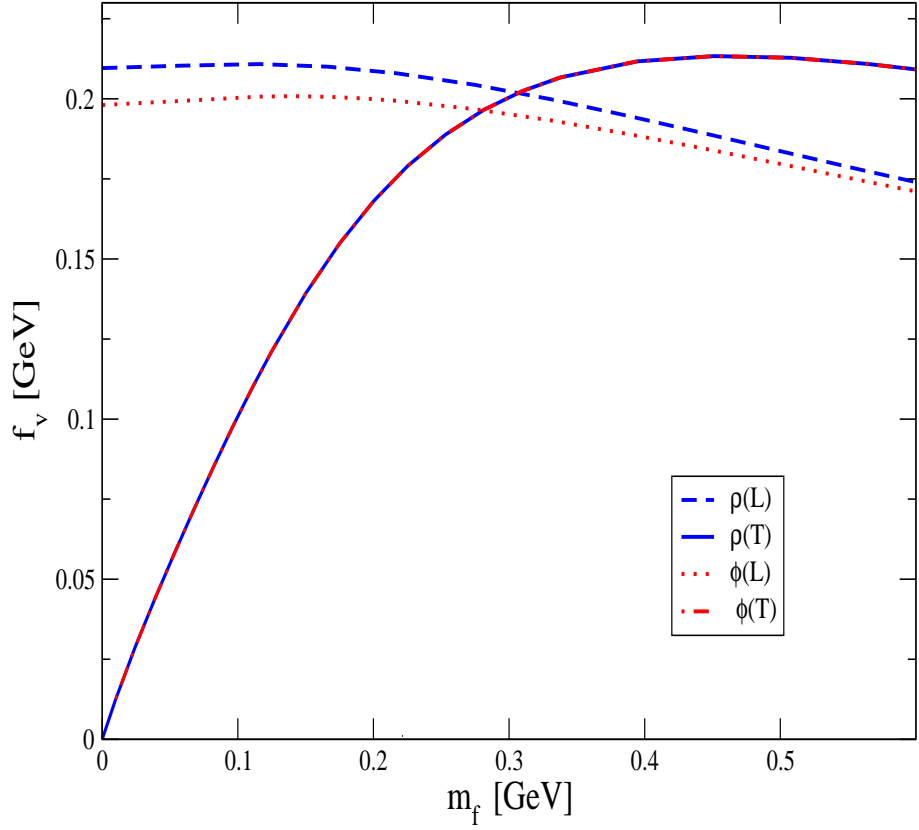}
\caption{The variation of the vector and tensor couplings with the quark mass.  Solid blue curve: $f_{\rho}$. Dash-dot red curve: $f_{\phi}$. Dashed blue curve: $f_{\rho}^T$ at $\mu \sim 1$ GeV. Dotted red curve: $f_{\phi}^{T}$ at $\mu \sim 1$ GeV.}
\label{fv-mf}
\end{figure}

Having specified the holographic wavefunction for the vector mesons, we are now able to predict their vector and tensor couplings defined by \cite{Ball:1998sk}
\begin{equation}
\langle 0|\bar q(0)  \gamma^\mu q(0)|V
(P,\lambda)\rangle =f_V M_\rho e_\lambda^{\mu}
\label{fv-def}
\end{equation}
and
\begin{equation}
\langle 0|\bar q(0) [\gamma^\mu,\gamma^\nu] q(0)|V (P,\lambda)\rangle =2 f_V^{T}  (e^{\mu}_{\lambda} P^{\nu} - e^{\nu}_{\lambda} P^{\mu}) \;.
\label{fvT-def}
\end{equation}
respectively.  In Eqs. \eqref{fv-def} and \eqref{fvT-def}, $\bar{q}$ and $q$ are the antiquark and quark fields evaluated at the same spacetime point, $P^\mu$ and $e^{\mu}_{\lambda}$ are the momentum and polarization vectors of the vector meson. Inserting the Fock expansion of the meson states in the right-hand-side of Eq. \eqref{fv-def} and Eq. \eqref{fvT-def}, we obtain \cite{Ahmady:2012dy}
\bea
f_{V} &=&  {\sqrt \frac{N_c}{\pi} }  \int_0^1 {\mathrm d} x  \left[ 1 + { m_{f}^{2}-\nabla_{r}^{2} \over x (1-x) M^{2}_{V} } \right] \left. \Psi_L(\zeta, x) \right|_{r=0}
\label{fvL}
\eea
and
\begin{equation}
f_{V}^{T}(\mu) =\sqrt{\frac{N_c}{2\pi}} m_f \int_0^1 {\mathrm d} x \; \int {\mathrm d} r \; \mu J_1(\mu r)  \frac{\Psi_T(\zeta,x)}{x(1-x)}
\label{fvT}
\end{equation}
respectively.  Note that the tensor coupling is dependent on the scale $\mu$ although we find that our predictions for $f_V^{T}(\mu)$ hardly depend on $\mu$ for $\mu \ge 1~\text{GeV}$. This means that our predictions are at some unspecified low scale $\mu \sim 1~\text{GeV}$. As is obvious from Eq. \eqref{fvT}, the tensor coupling vanishes as $m_f \to 0$, consistent with the requirement that the tensor current vanishes in the chiral limit. There is no such requirement for the vector current and indeed we predict a non-vanishing value for the vector coupling as $m_f \to 0$. We show the variation of the vector and tensor couplings with the quark mass in Figure \ref{fv-mf}. It is interesting to note that the vector coupling is maximum for $m_{u,d,s} \approx 0.140$ GeV. 

The vector coupling is also referred to as the decay constant as it is related to the measured electronic decay width $\Gamma_{V \rightarrow e^+ e^-}$ of the vector meson:
\be \Gamma_{V \rightarrow e^+ e^-}={ 4 \pi  \alpha_{em}^2  C_V^2 \over 3 M_V }f_V^2  
\label{width-fv}
\ee
 where $C_\phi=1/3$ and $C_\rho=1/\sqrt{2}$. Our results for the electronic decay widths are shown in Table \ref{tab:Decay-width}. Note that we obtain a lower value for the decay width of the $\rho$ than that reported in Ref. \cite{Forshaw:2012im} because we are using  a universal $\kappa=0.54$ GeV for both vector mesons compared to $\kappa=M_{\rho}/\sqrt{2}=0.55$ GeV used in Ref. \cite{Forshaw:2012im}. We show predictions for the decay width using $m_{u,d}=[0.046, 0.14]$ GeV for the $\rho$ and $m_{s}=[0.14, 0.357]$ GeV for the $\phi$. As can be seen in Figure \ref{fv-mf}, the vector coupling for the $\rho$ meson varies slowly with the quark mass in the range $m_f \in [0.046,0.14]$ GeV and hence our two predictions for the decay width do not differ much from each other with a slight preference for $m_{u,d}=0.14$ GeV. The variation of $f_{\phi}$ in the range $m_s \in [0.14,0.357]$ GeV is more important and the lower strange quark mass, $m_s=0.14$ GeV gives better agreement with the decay width datum.

 For both vector mesons, we underestimate the electronic decay width. But this is also the case with the other non-perturbative methods quoted in Table \ref{tab:fv-rho} for the $\rho$. There are likely perturbative corrections that must be taken into account when predicting the electronic decay width. 
 
\begin{table}
  \centering
  \begin{tabular}{c|c|c|c}
    \hline\hline
      Meson  &  $f_V$ [GeV] &$\Gamma_{e^+e^-}$ [KeV] & $\Gamma_{e^+e^-}$[KeV] (PDG) \\ \hline
      $\rho$ & $[0.210,0.211]$ & $[6.355,6.383]$ & $7.04 \pm 0.06$\\ \hline
      $\phi$ & $[0.205,0.191]$  & $[0.981,0.891]$  & $1.251 \pm 0.021$ \\ \hline
       \end{tabular}
  \caption{Predictions for the electronic decay widths of the $\rho$ and $\phi$ vector mesons using the holographic wavefunction given by Eq. \eqref{hwf} with $m_{u,d}=[0.046,0.14]$ GeV and $m_s=[0.14,0.357]$ GeV respectively.}
  \label{tab:Decay-width}
\end{table}

Since the optimal agreement (or rather minimal disagreement) with the electronic decay width data is achieved with $m_{u,d,s} \approx 0.14$ GeV, we choose this quark mass to compare our predictions with QCD Sum Rules, Dyson-Schwinger and lattice predictions as shown in Tables \ref{tab:fv-rho} and \ref{tab:fv-phi}. Recall that our predictions for the transverse decay constant are at $\mu \sim 1$ GeV which prevents an exact comparison with the other predictions all given at a scale $\mu=2$ GeV. Despite this, it is clear that we predict a smaller transverse decay constant (with $m_{u,d,s}=0.14$ GeV) than those predicted by the other non-perturbative methods quoted in Table \ref{tab:fv-rho}.

\begin{table}
  \centering
  \begin{tabular}{c|c|c|c|c}
    \hline\hline
    Reference & Approach & $f_{\rho}$ [MeV] & $f_{\rho}^{\perp}$ [MeV] & $f_{\rho}^{\perp}/f_{\rho}$\\ \hline
    This paper & LF holography & $211$  & $95$ & $0.45$\\ \hline
    Ref. \cite{Ball:1998kk} & Sum Rules & $198 \pm 7$ & $152 \pm 9$ &   \\ \hline
    Ref. \cite{Ball:2007zt} & Sum Rules & $206 \pm 7$ & $145 \pm 9$ &  $0.70 \pm 0.04$ \\ \hline
     Ref. \cite{Becirevic:2003pn} & Lattice (continuum) &  &  & $0.72 \pm 0.02$ \\ \hline
     Ref. \cite{Braun:2003jg} & Lattice (finite) &  &  & $0.742 \pm 0.014$ \\ \hline
     Ref. \cite{Jansen:2009hr} & Lattice (unquenched) &  & $159 \pm 0.008$ & $0.76 \pm 0.04$ \\ \hline
    Ref. \cite{Gao:2014bca} & Dyson-Schwinger  & $212$ & $156$ & $0.73$ \\ \hline

\end{tabular}
  \caption{Our predictions for the longitudinal and transverse decay constants and their ratio for the $\rho$ meson using $m_{u,d}=0.14$ GeV. Our predictions  for the transverse decay constant (and thus the ratio) is at a scale $\mu \sim 1$ GeV while the other non-perturbative predictions are at a scale $\mu=2$ GeV except for Ref. \cite{Ball:1998kk} where $\mu=2.2$ GeV. Note that we have multiplied the predictions of Ref. \cite{Gao:2014bca} by $\sqrt{2}$ to compare with our predictions.}
  \label{tab:fv-rho}
\end{table}

\begin{table}
  \centering
  \begin{tabular}{c|c|c|c|c}
    \hline\hline
    Reference & Approach & $f_{\phi}$ [MeV] & $f_{\phi}^{\perp}$ [MeV] & $f_{\phi}^{\perp}/f_{\phi}$\\ \hline
    This paper & LF holography & $201$  & $95$ & $0.48$\\ \hline
    Ref. \cite{Ball:1998kk} & Sum Rules & $254 \pm 3$ & $204 \pm 14$ &   \\ \hline
     Ref. \cite{Becirevic:2003pn} & Lattice (continuum) &  &  & $0.76 \pm 0.01$ \\ \hline
     Ref. \cite{Braun:2003jg} & Lattice (finite) &  &  & $0.780 \pm 0.008$ \\ \hline
     Ref. \cite{Jansen:2009hr} & Lattice (unquenched) &  & $$ & $$ \\ \hline

   Ref. \cite{Gao:2014bca} & Dyson-Schwinger & $190$ & $150$ & $0.79$ \\ \hline
\end{tabular}
  \caption{Our predictions for the longitudinal and transverse decay constants and their ratio for the $\phi$ meson using $m_{s}=0.14$ GeV. Our predictions  for the transverse decay constant (and thus the ratio) is at a scale $\mu \sim 1$ GeV while the other non-perturbative predictions are at a scale $\mu=2$ GeV except for Ref. \cite{Ball:1998kk} where $\mu=2.2$ GeV.}
  \label{tab:fv-phi}
\end{table}

\section{Refitting the CGC dipole model}
\label{F2 fit}

 In principle, the dipole-proton scattering amplitude $\mathcal{N}(x_{\text{m}},r,b)$ can be obtained by solving the Balitsky-Kovchegov (BK) equation \cite{Balitsky:1995ub,Kovchegov:1999yj,Kovchegov:1999ua} which itself can be derived within the Colour Glass Condensate (CGC) formalism \cite{JalilianMarian:1997jx,JalilianMarian:1997gr,Iancu:2000hn,Iancu:2001ad,Weigert:2000gi}. However, work is still in progress to implement in a satisfactory way the impact-parameter dependence in the proton-dipole amplitude \cite{GolecBiernat:2003ym,Berger:2010sh,Berger:2011ew}.  A widely used model is that proposed by Kowalski and Watt \cite{Watt:2007nr} where the saturation scale (see below) have a Gaussian dependence on the impact parameter. However, it has been recently argued in Ref. \cite{Contreras:2015joa} that the $b$-dependence should be exponential. In any case, considering the $b$-dependence introduces an additional parameter which has to be fixed using  data on diffractive meson production (for instance $J/\Psi$ production) which requires a model for the meson wavefunction. On the other hand, a simple model for the $b$-integrated dipole-proton amplitude, i.e. the dipole cross-section has been proposed long ago in Ref. \cite{Iancu:2003ge}. This is known as the CGC dipole model and is given by
\be 
\hat{\sigma}(x_{\text{m}},r) = \sigma_0 \, {{ \mathcal N } (x_{\text{m}}, r Q_s, 0 ) }\,.
\ee
with \bea
{ \mathcal N } (x_{\text{m}},r Q_s, 0) &=& { \mathcal N}_0 \left ( { r Q_s \over 2 }\right)^ {2 \left [ \gamma_s + { {\mathrm ln}  (2 / r Q_s) \over  \kappa \, \lambda \, {\mathrm ln} (1/x_{\text{m}}) }\right]}  ~~~~~~~{\rm for } ~~~~~ ~~~~ r Q_s \leq 2 \nonumber \\
&=& { 1- \exp[-{\mathcal A} \,{\mathrm ln}^2 ( {\mathcal B} \, r Q_s)]}  ~~~~~~~~~{\rm for } ~~~~~ ~~~~ r Q_s > 2
\eea
where the saturation scale $Q_s = (x_0/x_{\text{m}})^{\lambda / 2}$ GeV. The coefficients ${\mathcal A}$  and ${\mathcal B}$ are determined  from the condition that the ${\mathcal N}(r Q_s, x)$ and its derivative with respect to $r Q_s$ are  continuous at $r Q_s=2$.  This leads to
\be  {\mathcal A} =  - { ({ \mathcal N}_0 \gamma_s)^2  \over (1 - { \mathcal N}_0)^2 \, {\mathrm ln}[1 - { \mathcal N}_0] }\,, ~~~~~~
{\mathcal B} = {1 \over 2} (1 - { \mathcal N}_0)^{-{(1 - { \mathcal N}_0) \over { \mathcal N}_0 \gamma_s }}\,.
\ee
The free parameters of the CGC dipole model are $\sigma_0, \lambda, x_0$ and $\gamma_s$ which are fixed by a fit to the structure function $F_2$ data.
${ \mathcal N}_0$ and $\kappa$ are fixed at 0.7 and 9.9 (LO BFKL prediction) respectively. The high quality DIS data from HERA can be used to fix the free parameters of the dipole cross-section. An earlier fit to the structure function data given in Ref.\cite{Watt:2007nr} and used in Ref. \cite{Forshaw:2012im} to make successful predictions for diffractive $\rho$ production are: $\sigma_0 = 27.4$ mb, $\gamma_s = 0.74,  \lambda = 0.216, x_0 = 1.63 \times10^{-5}$.

In 2015, the H1 and ZEUS collaborations have released highly precise combined data sets \cite{Abramowicz:2015mha} for the reduced cross-section 
\begin{equation}
	\sigma_r(Q^2,x,y)=F_2(Q^2,x)-\frac{y^2}{1+(1-y)^2} F_L(Q^2,x)
	\label{sigmar}
\end{equation}
where $y=Q^2/\hat{s}x$ and $\sqrt{\hat{s}}$ is the centre of mass energy of the $e p$ system for $4$ different bins : $\sqrt{\hat{s}}=225$ GeV ($78$ data points),  $\sqrt{\hat{s}}=251$ GeV ($118$ data points) and $\sqrt{\hat{s}}=300$ GeV ($71$ data points), $\sqrt{\hat{s}}=318$ GeV ($245$ data points). The  structure functions in Eq. \eqref{sigmar} are given by
\begin{equation}
	F_2(Q^2,x_{\text{Bj}})=\frac{Q^2}{4\pi^2 \alpha_{\mbox{em}}} (\sigma_L^{\gamma^* p} (Q^2,x_{\text{Bj}}) + \sigma_T^{\gamma^* p}(Q^2,x_{\text{Bj}}))
\end{equation} 
and 
\begin{equation}
	F_L(Q^2,x_{\text{Bj}})= \frac{Q^2}{4\pi^2 \alpha_{\mbox{em}}} \sigma_L^{\gamma^* p} (Q^2,x_{\text{Bj}})
\end{equation}
where in the dipole model, $\sigma_{L,T}^{\gamma^* p}(Q^2,x_{\text{Bj}})$ is given by equation \eqref{gammapxsec}.
 
The most recent extraction of the CGC dipole model parameters was performed in Ref. \cite{Rezaeian:2013tka} following the release of the combined HERA data in 2010 \cite{Aaron:2009aa}. In this paper, the authors report a successful fit in the kinematic range $Q^2 \in [0.25,45]~\mbox{GeV}^2, x_{\text{Bj}} < 0.01$ using very small light quark masses $m_{u,d,s} \sim 10^{-3}$ GeV and a charm quark mass $m_c=1.27$ GeV. The fitted parameters are $\sigma_0 = 21.85$ mb, $\gamma_s = 0.762,  \lambda = 0.232, x_0 = 6.226 \times10^{-5}$ with a $\chi^2/\mbox{d.o.f}=1.18$. 
 
 We start by computing the $\chi^2$ per data point ($\chi^2/\text{d.p}$) for the 2015 HERA data using the earlier fitted parameters of Ref. \cite{Watt:2007nr} and the most recent fitted parameters of Ref. \cite{Rezaeian:2013tka}. We obtain $\chi^2/\mbox{d.p}=980/524=1.87$ and $\chi^2/\mbox{d.p}=562/524=1.07$ respectively. The latter $\chi^2/\mbox{d.p}$ is acceptable and in fact lower than that reported in Ref. \cite{Rezaeian:2013tka} for the fit to the 2010 data set. However, since we are using here effective quark masses and not current quark masses as in Ref. \cite{Rezaeian:2013tka}, we need to refit the CGC dipole parameters to the new 2015 data. To obtain the above $\chi^2/\mbox{d.p}$ values, we have used $m_c=1.27$ GeV as in Ref. \cite{Rezaeian:2013tka}, and we shall also use this same charm mass for our fits. Departing from Ref. \cite{Rezaeian:2013tka},  we also include low $Q^2 \in [0.045,0.25]~\mbox{GeV}^2$ data in our fits, thereby daring to extrapolate the use of the CGC dipole model in the non-perturbative region where the predictions are more sensitive to the quark masses.

   Our fitted values for the CGC dipole model parameters  together with the resulting  $\chi^2$ per degrees of freedom ($\chi^2/\mbox{d.o.f}$) values are shown in Table \ref{tab:F2fit}. The first two rows indicate that the fit is not very sensitive to the variation in the strange quark mass. Comparing the second and third rows, we can see that the data prefer the lower $u$ and $d$ quark masses and that increasing them give quite different fit parameters especially for $x_0$. But in all three cases, the  $\chi^2/\mbox{d.o.f}$ is less than the value obtained in Ref. \cite{Rezaeian:2013tka} ($\chi^2/\mbox{d.o.f}=1.18$) and we regard them as acceptable fits. This is not to deny that using current quark masses can lead to equally good fits. In fact, we find that with current quark masses $m_{u,d,s}= 0.001$ GeV, we obtain $\chi^2/\text{d.o.f}=528/520=1.01$, i.e. a fit of similar quality as our best fit with $m_{u,d}=0.046$ GeV and $m_s=0.14$ GeV. The fitted parameters are $\gamma_s=0.76$, $\lambda=0.236$, $x_0=6.2 \times 10^{-5}$ and $\sigma_0=21.6$ mb, similar to those reported in Ref. \cite{Rezaeian:2013tka}. However, we have checked that this quark mass set (with our holographic wavefunction) does not lead to a good agreement with the diffractive cross-section data at low $Q^2$.

\begin{table}
  \centering
  \begin{tabular}{c|c|c|c|c|c}
    \hline\hline
      $[m_{u,d},m_s]$/GeV  &  $\gamma_s$  &$\sigma_0$/mb & $x_0$ & $\lambda$ & $\chi^2/\mbox{d.o.f}$ \\ \hline
 $ [0.046,0.357]$ & $0.741$  &  $26.3$ & $1.81 \times 10^{-5}$ & $0.219$ & 535/520=1.03\\ \hline
 $ [0.046,0.14]$ & $0.722$  &  $24.9$ & $1.80 \times 10^{-5}$ & $0.222$ & 529/520=1.02\\ \hline
$ [0.14,0.14]$ & $0.724$  &  $29.9$ & $6.33 \times 10^{-6}$ & $0.206$ & 554/520=1.07 \\ \hline

       \end{tabular}
  \caption{Parameters of the CGC dipole model extracted from our fits to inclusive DIS data (with $x_{\text{Bj}} \le 0.01$ and $Q^2\in [0.045, 45]\,\text{GeV}^2$) using $3$ different set of quark masses. The error on each parameter is less than $1\%$, i.e changing one parameter by its error, increases the $\chi^2/\mbox{d.o.f}$ by less than $1\%$.}
  \label{tab:F2fit}
\end{table}

\section{Predicting diffractive cross-sections}
\label{Diffractive data}
Having specified the dipole cross-section and the holographic meson wavefunction, we can now compute cross-sections for diffractive $\rho$ and $\phi$ production. We shall show predictions using three sets of the CGC dipole parameters as given in Table \ref{tab:F2fit}. We shall refer to these three sets of predictions as "Fit A" (first row), "Fit B" (second row) and "Fit C" (third row) respectively. Recall that all our predictions will be generated using the same holographic wavefunction given by Eq. \eqref{hwf} and they differ only by the choice of quark masses and the corresponding fitted parameters of the CGC dipole model as given in Table \ref{tab:F2fit}. 

We compute the total cross-section as a function of $W$ in different $Q^2$ bins as well as a function of $Q^2$ at fixed $W$. We also compute the ratio of longitudinal to transverse cross-sections as a function of $Q^2$ at fixed $W$. Predicting the latter observable is interesting since the normalization uncertainties in the diffractive $B$-slope and the dipole cross-section, cancel out, increasing its sensitivity to the meson wavefunction.

\begin{figure}[htbp]
\centering 
\includegraphics[width=14cm,height=14cm]{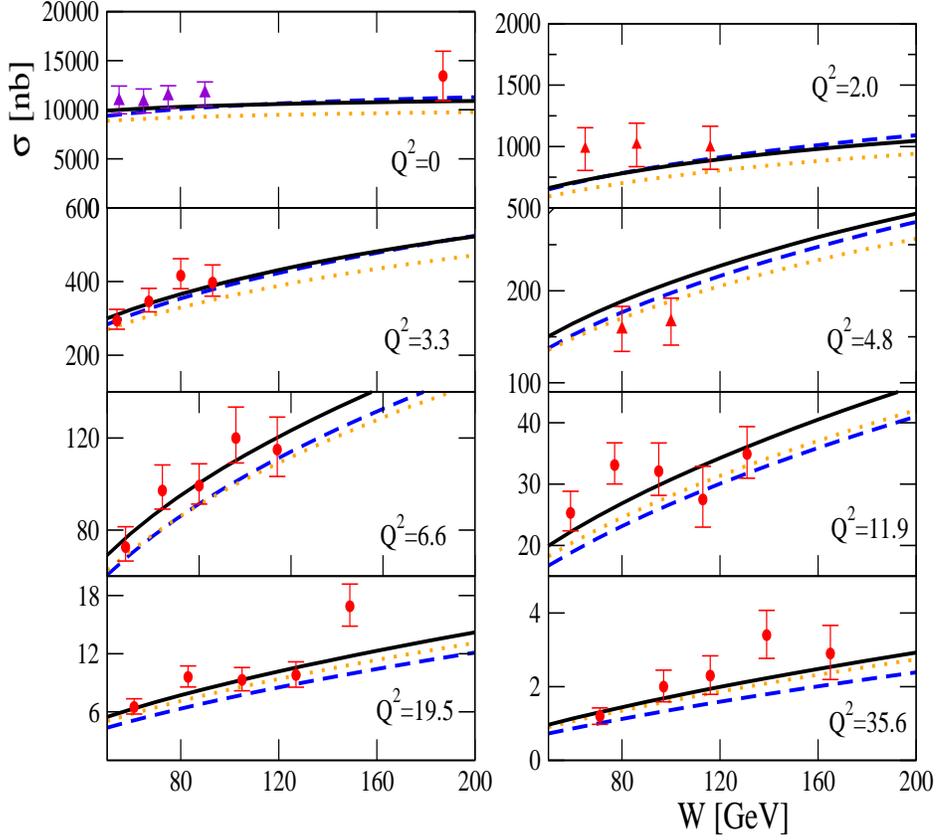}
\caption{Predictions for $\rho$ production cross section as a function of $W$ in different $Q^2$ bins compared to H1 data \cite{Aid:1996bs,Aaron:2009xp} and ZEUS data (at $Q^2=0$) \cite{Breitweg:1997ed}. Black solid curves: Fit A. Orange dotted curves: Fit B. Blue dashed curves: Fit C.}
\label{rho:sigmaW-H1}
\end{figure}

\begin{figure}[htbp]
\centering 
\includegraphics[width=14cm,height=14cm]{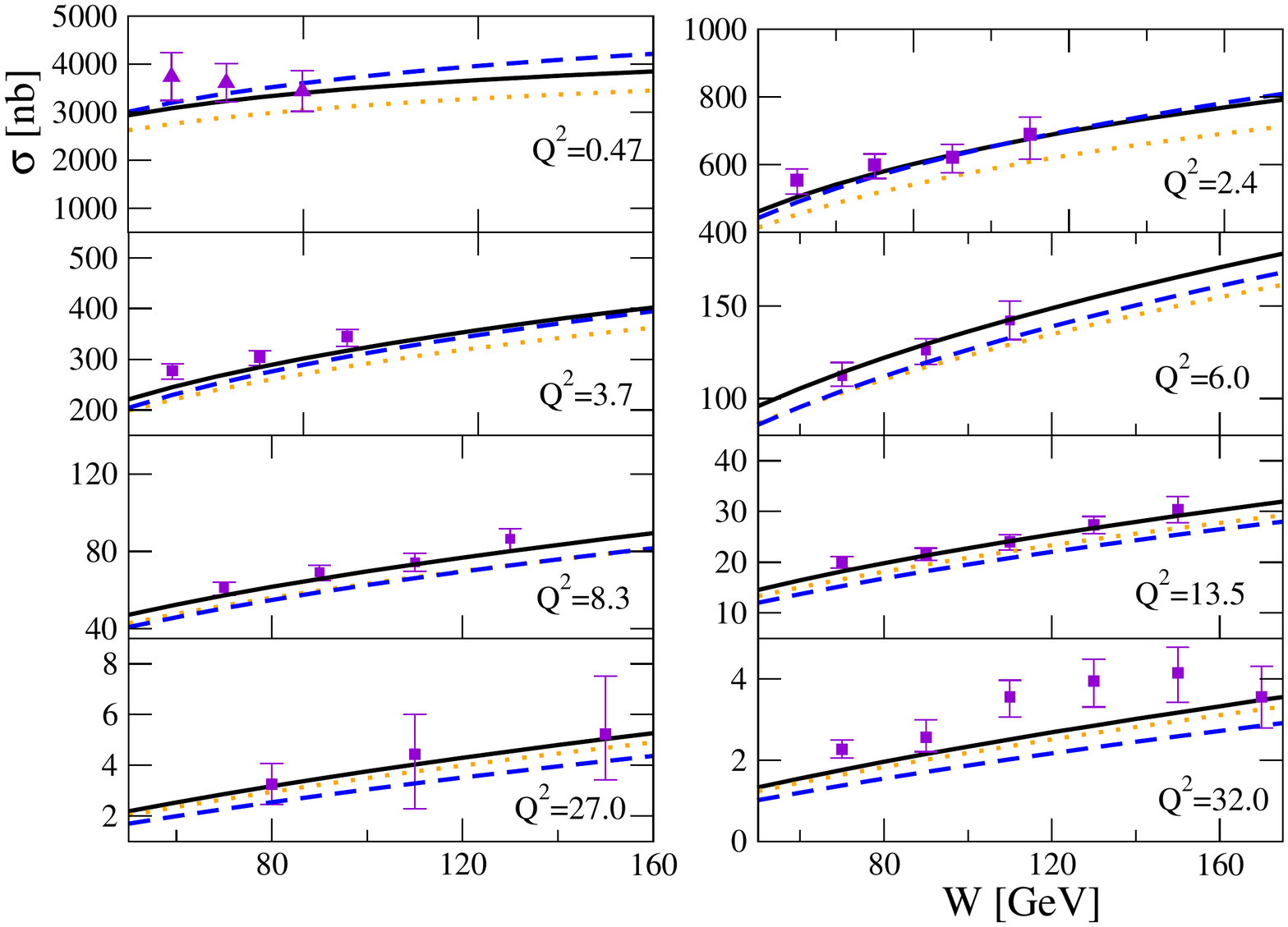}
\caption{Predictions for $\rho$ production cross section as a function of $W$ in different $Q^2$ bins compared to ZEUS data \cite{Aaron:2009xp}. Black solid curves: Fit A. Orange dotted curves: Fit B. Blue dashed curves: Fit C.}
\label{rho:sigmaW-ZEUS}
\end{figure}

\begin{figure}[htbp]
\centering 
\includegraphics[width=14cm,height=14cm]{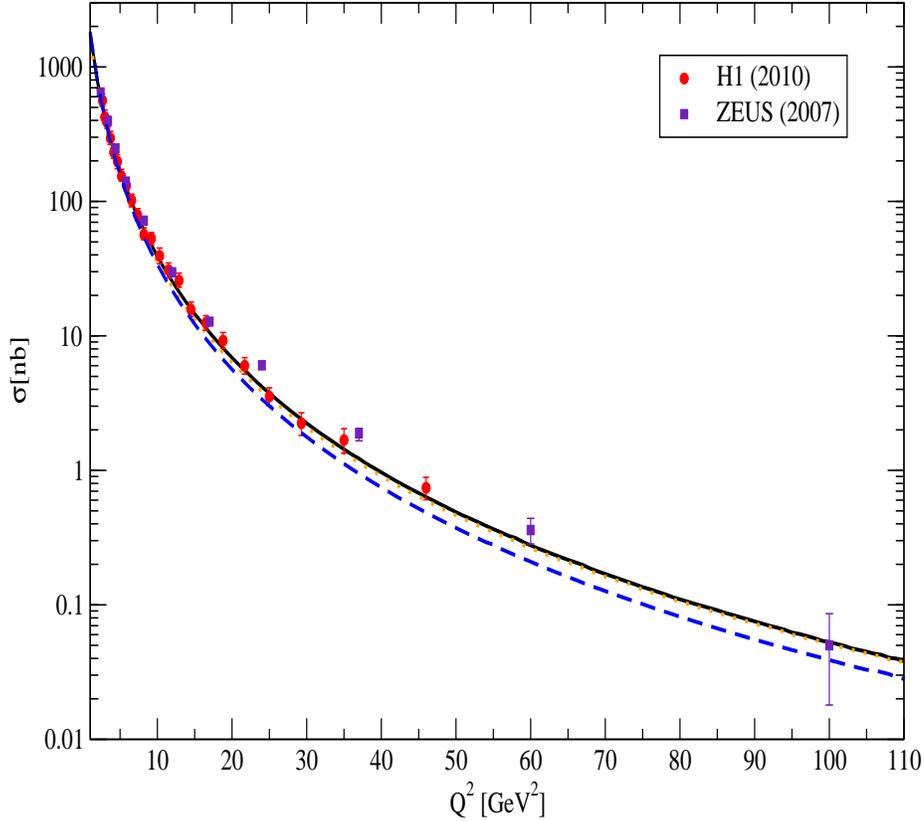}
\caption{Predictions for $\rho$ production cross section as a function of $Q^2$ at $W=75$ GeV, compared to HERA data \cite{Aaron:2009xp}.  Black solid curve: Fit A. Orange dotted curve: Fit B. Blue dashed curve: Fit C.}
\label{rho:sigmaq2}
\end{figure}

\begin{figure}[htbp]
\centering 
\includegraphics[width=12cm,height=12cm]{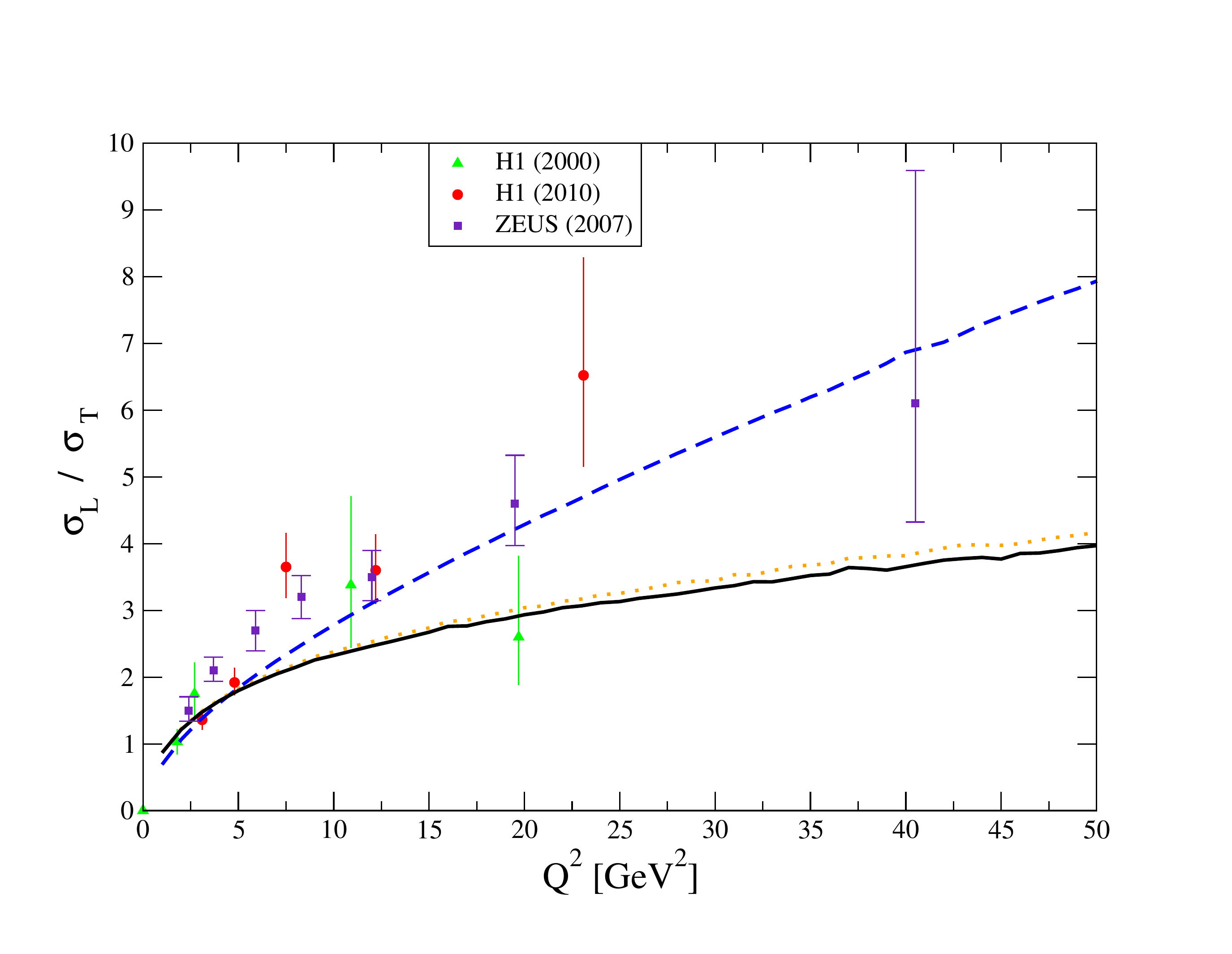}
\caption{Predictions for the $\rho$ production longitudinal to transverse cross-section ratio in $\rho$ production as a function of $Q^2$ at $W=75$ GeV compared to the H1 data (at $W=75$ GeV) \cite{Aaron:2009xp,Adloff:1999kg} and ZEUS data (at $W=90$ GeV) \cite{Chekanov:2007zr}. Black solid curve: Fit A. Orange dotted curve: Fit B. Blue dashed curve: Fit C. }
\label{rho:Ratio}
\end{figure}

For $\rho$ production, our predictions  for the $W$ dependence of the total cross-section in different $Q^2$ bins are shown in Figures \ref{rho:sigmaW-H1} and \ref{rho:sigmaW-ZEUS} while our predictions for the $Q^2$ dependence of the total cross-section at fixed $W$ are shown in Figure \ref{rho:sigmaq2}. The "Fit A" (black solid curves) and the "Fit C" (blue dashed curves) are both accommodated by the total cross-section data.  The $\sigma_L/\sigma_T$ ratio data, shown in Figure \ref{rho:Ratio}, are able to discriminate between them and favour the "Fit C" prediction. Notice that the "Fit C" predictions  undershoot the data in the two largest $Q^2$ bins but this is the kinematic region where the non-perturbative holographic wavefunction is expected to be less accurate.  We thus confirm the conclusion of Ref. \cite{Forshaw:2012im} in which the set of equal quark masses was used.

\begin{figure}[htbp]
\centering 
\includegraphics[width=12cm,height=12cm]{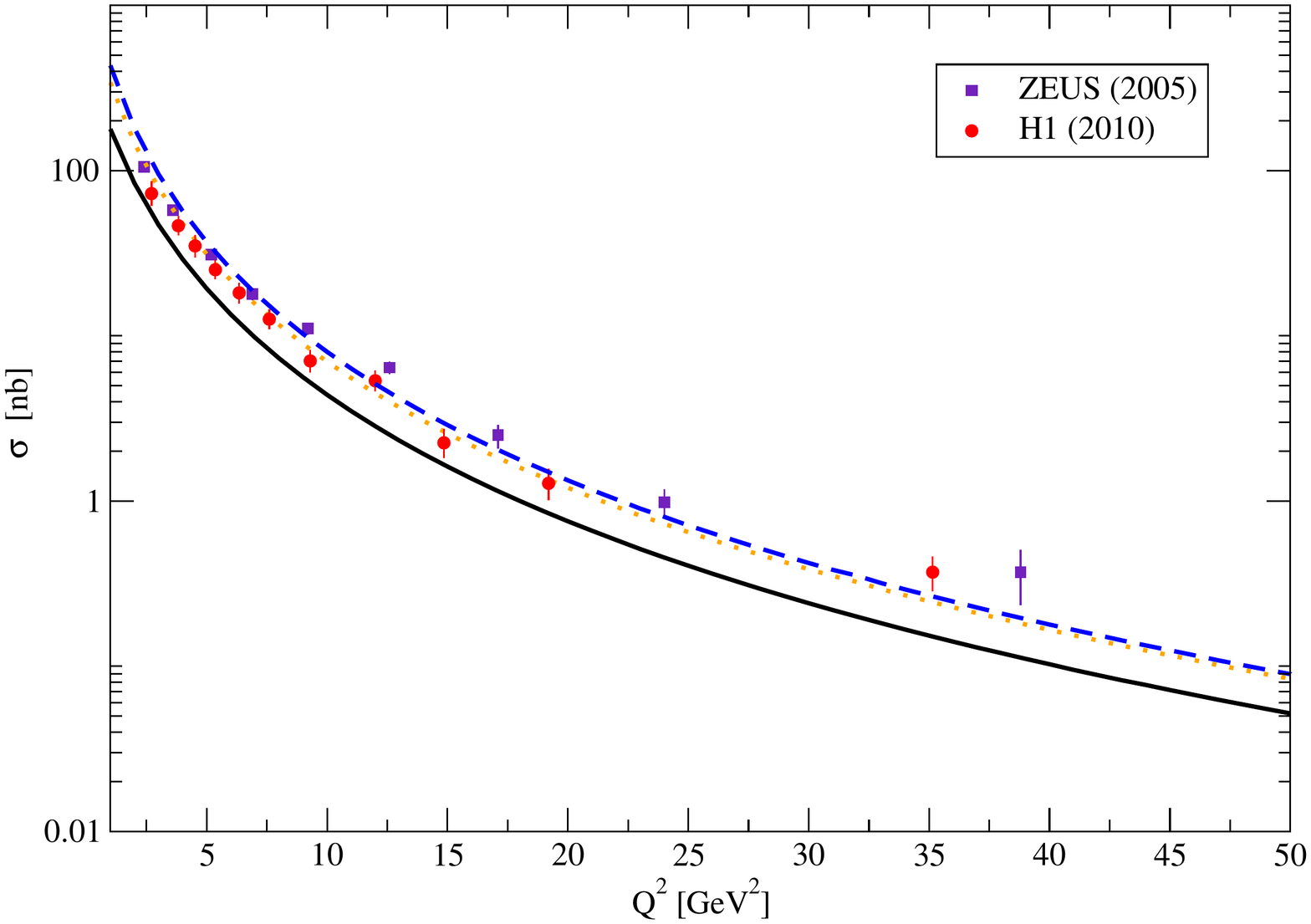}
\caption{ Predictions for the $\phi$ production total cross section for at $W=90$ GeV as a function of $Q^2$  compared to HERA data \cite{Chekanov:2005cqa,Aaron:2009xp}. Black solid curve: Fit A. Orange dotted curve: Fit B. Blue dashed curve: Fit C.}
\label{phi-q2}
\end{figure}

\begin{figure}[htbp]
\centering 
\includegraphics[width=14cm,height=14cm]{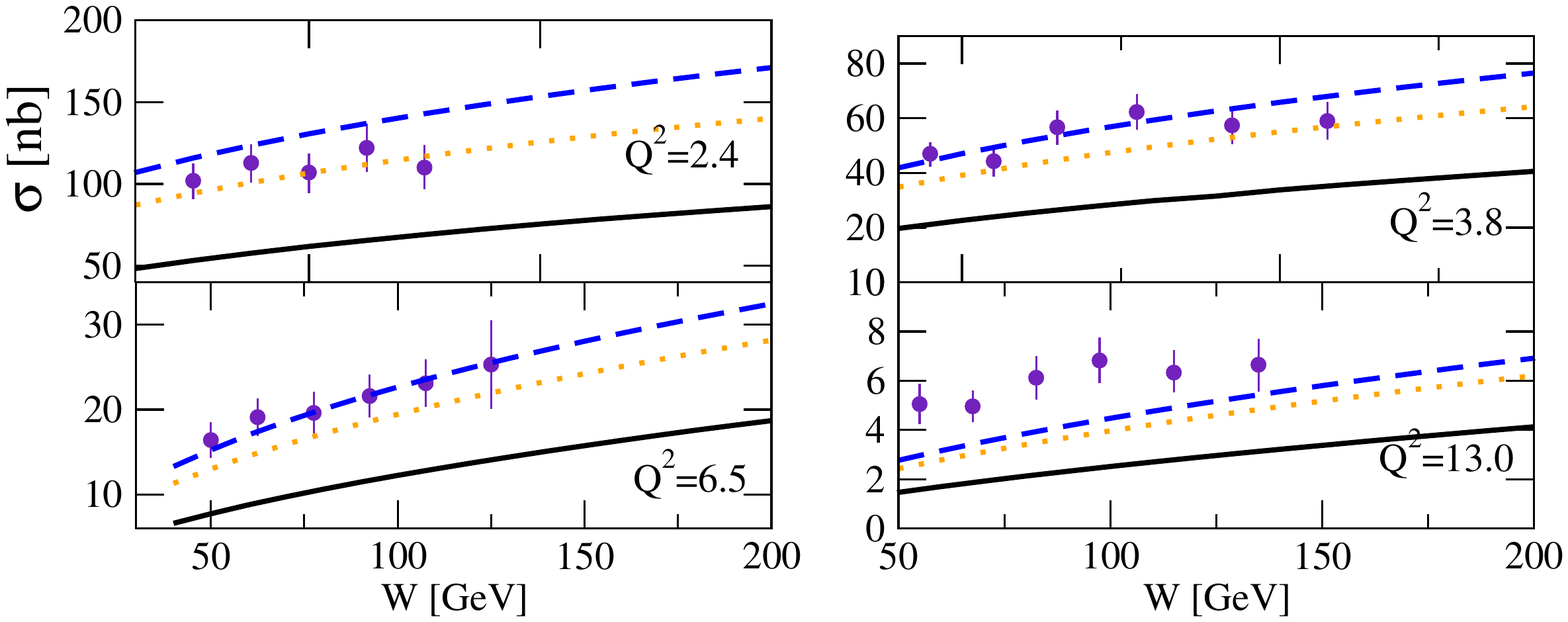}
\caption{Our predictions for $\phi$ production total cross-section for $\gamma^* p \to \phi p$ as a function of $W$ in different $Q^2$ bins compared to the ZEUS data \cite{Chekanov:2005cqa}.  Black solid curves: Fit A. Orange dotted curves: Fit B. Blue dashed curves: Fit C.} 
\label{phi-W-ZEUS}
\end{figure}

\begin{figure}[htbp]
\centering 
\includegraphics[width=14cm,height=14cm]{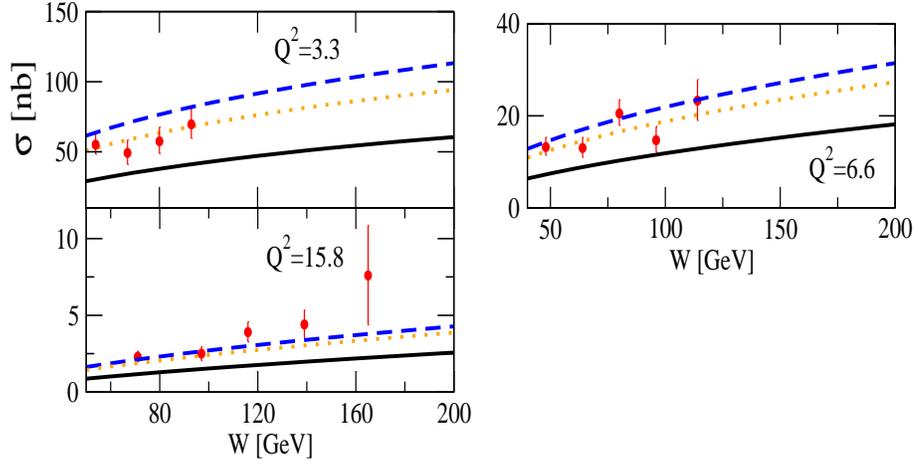}
\caption{Our predictions for the $\phi$ production total cross-section as a function of $W$ in different $Q^2$ bins compared to the H1 data \cite{Aaron:2009xp}.Black solid curves: Fit A. Orange dotted curves: Fit B. Blue dashed curves: Fit C.} 
\label{phi-W-H1}
\end{figure}

\begin{figure}[htbp]
\centering 
\includegraphics[width=14cm,height=14cm]{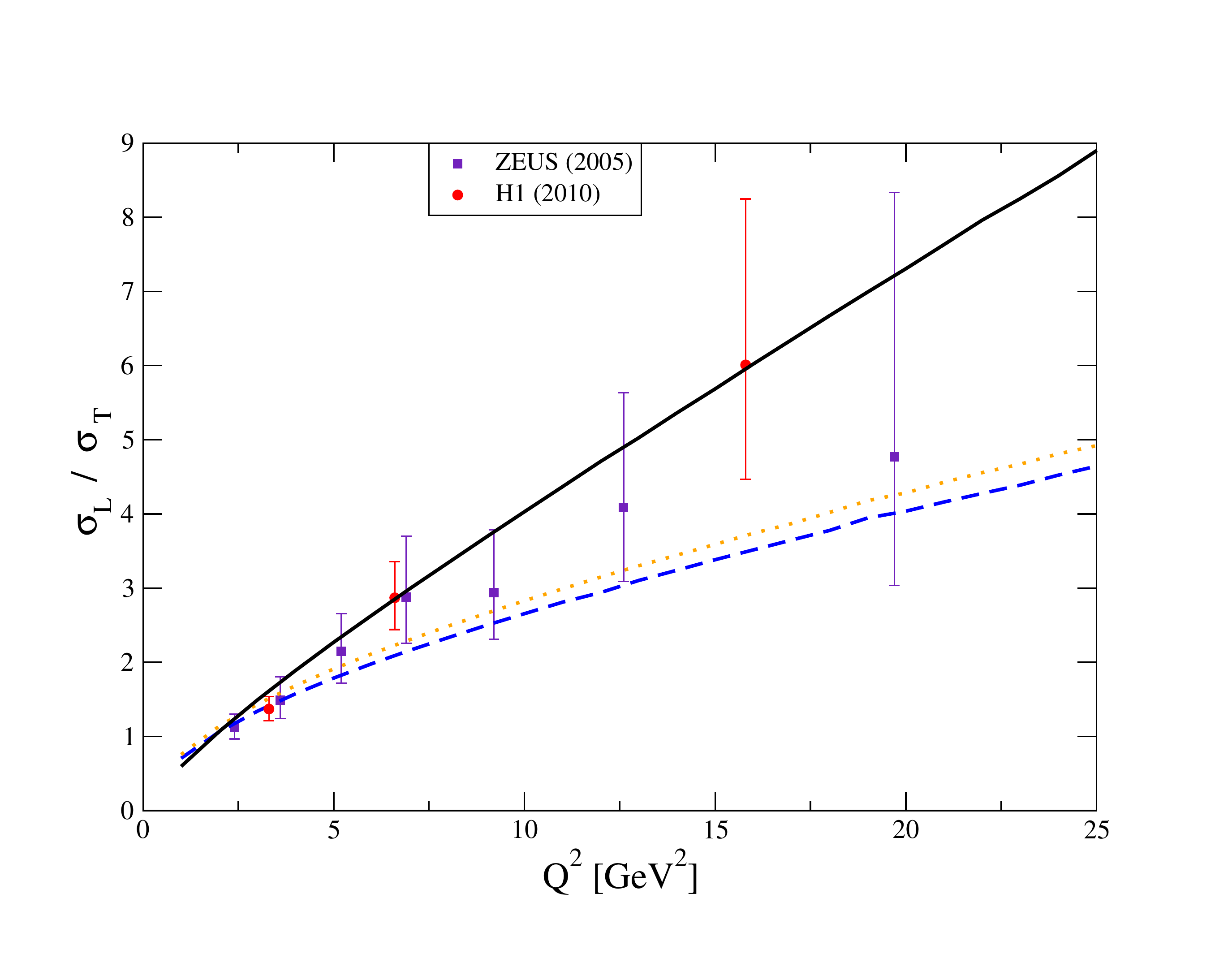}
\caption{Our predictions for the $\phi$ production longitudinal to transverse cross-section ratio at $W=90$ GeV compared to the ZEUS data \cite{Chekanov:2005cqa}. Black solid curve: Fit A. Red dotted curves: Fit B. Blue dashed curve: Fit C.} 
\label{phi-ratio}
\end{figure}

For $\phi$ production, our predictions for the $W$ dependence of the total cross-section  in different $Q^2$ bins are shown in Figures \ref{phi-W-ZEUS} and \ref{phi-W-H1} and for the $Q^2$ dependence of the total cross-section at fixed $W$ are shown in Figure \ref{phi-q2}. Here, it is clear that the "Fit A" predictions (solid black curves) are not successful. The data prefer slightly the "Fit B" (orange dotted curves) over the "Fit C" predictions (blue dashed curves) although the lack of data in the low $Q^2$ region prevents us from making a definite statement. At high $Q^2$, all our predictions tend to undershoot the (ZEUS) data as expected. Our predictions for the longitudinal to transverse cross-sections ratio for $\phi$ production are shown in Fig. \ref{phi-ratio}. We can see that the ratio data tend to favour the "Fit A"  prediction (solid black curve) although they are not precise enough to discard the other two predictions.   

In summary, the "Fit C" predictions are favoured for $\rho$ production and the "Fit B" predictions are preferred for $\phi$ production. Notice that in both sets of predictions, we are using exactly the same holographic wavefunction for both $\rho$ and $\phi$, i.e. a wavefunction with the same $\kappa=0.54$ GeV and $m_{f}=0.14$ GeV.  The difference between the two sets of predictions arises from the different electromagnetic couplings and the different fitted parameters of the CGC dipole cross-section. 

\begin{figure}[htbp]
\centering 
\includegraphics[width=14cm,height=14cm]{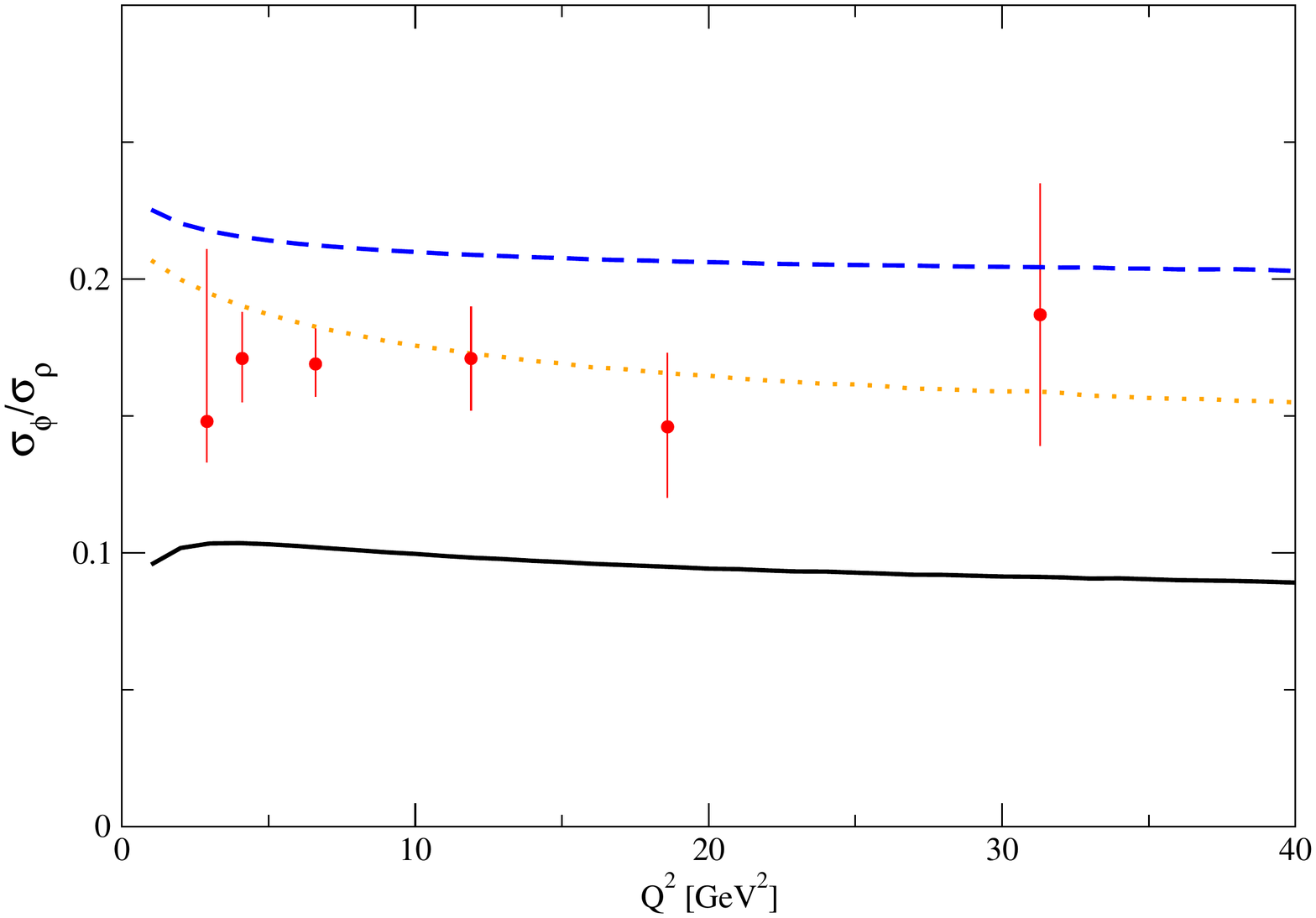}
\caption{Our predictions for the ratio of the total cross-sections for $\phi$ and $\rho$ production at $W=75$ GeV compared to the H1 data \cite{Aaron:2009xp}. Black solid curve: Fit A. Red dotted curves: Fit B. Blue dashed curve: Fit C.}
\label{phi-rho-R}
\end{figure}

Finally, we consider the data set on the ratio of the total cross-sections for $\phi$ and $\rho$ production. Note if the $\rho$ and the $\phi$ had identical masses and holographic wavefunctions, this ratio is simply given by the squared ratio of the effective electric charges of the quark-antiquark dipole coupling to the photon: $e_{s}^2/e_{u/d}^2=(1/3)^2/(1/\sqrt{2})^2= 0.22$. As expected our "Fit A" prediction (blue dashed curve) tends to that value at high $Q^2$. At lower $Q^2$, the deviation from $0.2$ is due to the vector meson mass entering the modified Bjorken-$x$ given by Eq. \eqref{Bjorken-x} and the diffractive slope given by Eq. \eqref{Bslope}. However, the data indicate a lower ratio and as can be seen in Figure \ref{phi-rho-R}, the "Fit B" prediction (orange dotted curve) is clearly preferred. This provides evidence for the need to have different quark masses in the holographic wavefunctions of the $\rho$ and $\phi$.

In view of the above results, we anticipate that a set of quark masses with an even weaker SU(3) symmetry breaking than our "Fit B" set, should give  the best simultaneous description of both $\rho$ and $\phi$ diffractive production data.

\section{Conclusion}
\label{Conclusion}
We have updated the parameters of the CGC dipole model using the definitive 2015 HERA data on inclusive DIS and we have used the fitted dipole cross-section together with a holographic meson wavefunction  in order to compute the cross-sections for diffractive $\rho$ and $\phi$ meson production. The holographic light-front meson wavefunction is successful to describe simultaneously diffractive $\rho$ and $\phi$ production with a single universal holographic mass scale $\kappa=0.54~\mbox{GeV}$ but with a set of light quark masses with a weaker SU(3) flavour symmetry ($m_s/m_{u,d} \lesssim 3$) breaking than that used in light-front holography ($m_s/m_{u,d} \approx 7$) to generate the pion and kaon masses. 

\section{Acknowledgements}
 This work of NS is supported by the Department of Science and Technology, Government of India, under the Fast Track scheme (Ref No. SR/FTP/PS-057/2012). The work of MA and RS is supported by a team grant (SAPGP-2014-00002) from the National Science and Engineering Research Council of Canada (NSERC). We thank S. J. Brodsky and J. R. Forshaw for their valuable comments as well as A. H. Rezaeian for useful correspondence.  
 
\bibliographystyle{apsrev}
\bibliography{revisedphi.bib}

\end{document}